\documentclass[11pt,a4paper]{article} 
\pdfoutput=1
\usepackage{jheppub,hyperref,mdwlist}
\usepackage{amsmath}
\usepackage{graphicx}
\usepackage{subfigure}
\usepackage{amssymb}
\usepackage{xcolor}
\usepackage{multirow}
\usepackage{cancel}
\usepackage{color}
\usepackage{listings}
\usepackage{verbatim}
\usepackage{float}
\usepackage{slashed}
\usepackage{mathtools}
\usepackage{soul}

\usepackage{tikz}
\usetikzlibrary{arrows,shapes}
\usetikzlibrary{trees}
\usetikzlibrary{matrix,arrows} 				
\usetikzlibrary{positioning}				
\usetikzlibrary{calc,through}				
\usetikzlibrary{decorations.pathreplacing}  
\usepackage{pgffor}							

\usetikzlibrary{decorations.pathmorphing}	
\usetikzlibrary{decorations.markings}
\tikzset{
    vector/.style={decorate, decoration={snake}, draw},
	provector/.style={decorate, decoration={snake,amplitude=2.5pt}, draw},
	antivector/.style={decorate, decoration={snake,amplitude=-2.5pt}, draw},
    fermion/.style={draw=black, postaction={decorate},
        decoration={markings,mark=at position .55 with {\arrow[draw=black]{>}}}},
    fermionbar/.style={draw=black, postaction={decorate},
        decoration={markings,mark=at position .55 with {\arrow[draw=black]{<}}}},
    fermionnoarrow/.style={draw=black},
    gluon/.style={decorate, draw=black,
        decoration={coil,amplitude=4pt, segment length=5pt}},
    scalar/.style={dashed,draw=black, postaction={decorate},
        decoration={markings,mark=at position .55 with {\arrow[draw=black]{>}}}},
    scalarbar/.style={dashed,draw=black, postaction={decorate},
        decoration={markings,mark=at position .55 with {\arrow[draw=black]{<}}}},
    scalarnoarrow/.style={dashed,draw=black},
    electron/.style={draw=black, postaction={decorate},
        decoration={markings,mark=at position .55 with {\arrow[draw=black]{>}}}},
	bigvector/.style={decorate, decoration={snake,amplitude=4pt}, draw},
}

\tikzstyle{block} = [draw, rectangle, 
    minimum height=3em, minimum width=6em]

\usepackage{amsmath}
\usepackage{wrapfig}
\usepackage{cleveref}

\newcommand{\be}{\begin{equation}}
\newcommand{\ee}{\end{equation}}
\newcommand{\beq}{\begin{equation}}
\newcommand{\eeq}{\end{equation}}
\newcommand{\bea}{\begin{eqnarray}}
\newcommand{\eea}{\end{eqnarray}}
\newcommand{\besp}{\begin{equation}\begin{split}}
\newcommand{\eesp}{\end{split}\end{equation}}

\newcommand{\nn}{\nonumber}

\newcommand{\Eq}[1]{Eq.~(\ref{#1})}

\newcommand{\Dfbd}{\mathord{\buildrel{\lower3pt\hbox{$\scriptscriptstyle\leftrightarrow$}}\over {D}_{\mu}}}
\newcommand{\ave}[1]{\left\langle #1\right\rangle}

\def\mD{\mathcal{D}}

\def\mL{\mathcal{L}}

\def\mO{\mathcal{O}}

\def\mS{\mathcal{S}}

\def\0{\textbf{0}}
\def\1{\textbf{1}}
\def\2{\textbf{2}}
\def\3{\textbf{3}}
\def\4{\textbf{4}}
\def\5{\textbf{5}}
\def\6{\textbf{6}}
\def\7{\textbf{7}}
\def\8{\textbf{8}}
\def\9{\textbf{9}}

\def\d{\text{d}}

\definecolor{RoyalBlue}{cmyk}{1, 0.50, 0, 0}

\begin{document}

\title{Dark matter in classically conformal theories: WIMP and supercooling}

\author[a]{Ke-Pan Xie}
\affiliation[a]{School of Physics, Beihang University, Beijing 100191, China}

\author[, a]{and Cheng-Hao Zhan\footnote{Corresponding author.}}

\emailAdd{kpxie@buaa.edu.cn}
\emailAdd{chzhan@buaa.edu.cn}

\abstract{
Beyond solving the hierarchy problem, classically conformal (CC) theories naturally accommodate dark matter (DM). In this work, we explore the CC $SU(2)_X$ gauge theory with a triplet dark scalar, uncovering two distinct DM scenarios: weakly interacting massive particle (WIMP) and supercooled DM. The production mechanisms are strongly influenced by the CC model's unique first-order phase transition evolution history, which differs significantly from those in non-conformal models. We obtain the viable parameter space for each scenario and investigate the current constraints and future sensitivities at experiments, demonstrating that gravitational wave signals from the phase transition provide a common detection channel for both the WIMP and supercooled DM regimes.
}

\maketitle
\flushbottom

\section{Introduction}

As the only elementary scalar in the Standard Model (SM), the Higgs boson suffers from the quadratic divergence in its self-energy from quantum corrections at high scales, known as the gauge hierarchy problem~\cite{Peskin:2025lsg}. A promising solution is the classically conformal (CC) principle~\cite{Bardeen:1995kv,Meissner:2007xv}, where the theory contains no dimensionful parameters at tree level, preserving conformal invariance. The conformal symmetry is then broken radiatively via the Coleman-Weinberg (CW) potential~\cite{Coleman:1973jx,Jackiw:1974cv}, dynamically generating the scale. Given the observed Higgs boson mass $m_h \approx 125~{\rm GeV}$~\cite{ATLAS:2024fkg}, a viable realization of this idea is to first trigger conformal symmetry breaking in a hidden sector through the CW mechanism, and then transmit it partially to the SM electroweak (EW) sector via Higgs portal interactions~\cite{Hempfling:1996ht,Iso:2009ss,Iso:2009nw,Chun:2013soa,Das:2015nwk,Jung:2021vap,deBoer:2024jne}.

Beyond the particle physics motivation, CC theories also offer solutions to puzzles in cosmology, such as dark matter (DM), which contributes $\sim26\%$ to the total energy of the Universe~\cite{Planck:2018vyg}. An appealing minimal DM scenario within this framework is a gauged hidden sector, where the new gauge boson plays a dual role: it generates a CW potential for the dark scalar to break the conformal and gauge symmetries, and it serves as a vector DM candidate. Previous studies have explored this scenario for a $U(1)_X$ group~\cite{YaserAyazi:2019caf,Mohamadnejad:2019vzg,Baules:2025pww,Frandsen:2022klh} and for an $SU(2)_X$ group with the dark scalar in the fundamental representation~\cite{Hambye:2018qjv,Baldes:2018emh,Borah:2021ftr,Kierkla:2022odc,Frandsen:2022klh}.

In this research, we investigate the DM scenario within a CC $SU(2)_X$ theory, featuring a dark scalar $\phi$ in the adjoint representation. The non-Abelian nature of the dark gauge group prevents any kinetic mixing with the SM hypercharge $U(1)_Y$, thereby ensuring the stability of the vector DM candidate. Moreover, a triplet $\phi$ induces the symmetry-breaking pattern $SU(2)_X \to U(1)_X$, and this residual unbroken group guarantees the absolute stability of vector DM. This is different from scenarios where the dark scalar is a doublet; in those models, $SU(2)_X$ is completely broken, and non-renormalizable operators can mediate DM decay to the SM particles~\cite{Hambye:2008bq}. An additional feature of the triplet scenario is the existence of a topological soliton, i.e., the 't Hooft-Polyakov monopole~\cite{tHooft:1974kcl,Polyakov:1974ek}, offering an extra DM candidate.

DM within non-conformal $SU(2)_X$ theories has been studied extensively, with the dark scalar in the fundamental~\cite{Hambye:2008bq,Baouche:2021wwa,Benincasa:2025tdr,Foguel:2025hio}, adjoint~\cite{Baek:2013dwa,Khoze:2014woa,Chaffey:2019fec,Ghosh:2020ipy,Nomura:2020zlm,Hu:2021pln}, or higher representations~\cite{Chen:2015nea,Zhang:2022wek,Yuan:2024tfs}. In those models, tree-level mass parameters trigger the $SU(2)_X$ and EW symmetries breaking, and the cosmological evolution typically follows a standard adiabatic thermal history, leading to freeze-out of the weakly interacting massive particle (WIMP) DM. The CC framework in the current paper differs in two crucial aspects:
\begin{enumerate}
\item Symmetry breaking in CC theories is driven by radiative corrections, resulting in a more predictive scenario with fewer free parameters. 
\item As first proposed by Witten~\cite{Witten:1980ez} and now actively studied~\cite{Konstandin:2011dr,Jinno:2016knw,Chiang:2017zbz,Iso:2017uuu,Marzo:2018nov,Bian:2019szo,Ellis:2019oqb,Ellis:2020nnr,Kawana:2022fum,Ahriche:2023jdq,Sagunski:2023ynd,Liu:2024fly,Salvio:2026bco}, CC theories usually feature a supercooled first-order phase transition (FOPT), which significantly impacts the cosmological history and formation of DM~\cite{Hambye:2018qjv,Baldes:2018emh,Kang:2020jeg,Khoze:2022nyt,Wong:2023qon,Ayazi:2025sfs,Athron:2025pog,Racco:2025ons} or primordial black holes (PBHs)~\cite{Baldes:2023rqv,Salvio:2023ynn,Gouttenoire:2023pxh,Arteaga:2024vde,Banerjee:2024cwv,Banerjee:2024fam,Cao:2025jwb},\footnote{Recently, PBH formation in slow FOPTs was challenged on gauge issues, as the comoving‑gauge curvature perturbation appears too small to form PBHs~\cite{Franciolini:2025ztf,Wang:2026zvz,Ai:2026zrs}. However, PBHs can still form if the post‑FOPT reheating is slow~\cite{Ai:2026zrs}.} as well as the generation of the baryon asymmetry~\cite{Baldes:2021vyz,Huang:2022vkf,Chun:2023ezg}.
\end{enumerate}

In this work, we identify two distinct DM scenarios within the CC $SU(2)_X$ model with a triplet scalar: (i) the dark gauge boson $X$ as a WIMP via the conventional freeze-out mechanism; and (ii) supercooled DM, where an ultra-supercooled FOPT leads to dilution and nonthermal production of $X$ after the transition~\cite{Hambye:2018qjv,Baldes:2018emh,Wong:2023qon,Ayazi:2025sfs}. We also examine the possibility of monopole DM arising from bubble nucleation dynamics, and obtain that it cannot dominate the DM abundance, consistent with Ref.~\cite{Brummer:2025inh}. For both viable scenarios, we identify the parameter space that explains the full DM relic abundance and explore their phenomenological signatures.

This paper is organized as follows. In Section~\ref{sec:model}, we introduce the model Lagrangian and derive the particle spectrum. Section~\ref{sec:history} then describes the unique cosmological thermal history in this model, which distinguishes the CC theory from non-conformal models. The two DM scenarios are detailed in Section~\ref{sec:dm}, while the DM parameter space and the corresponding phenomenological analysis are presented in Section~\ref{sec:pheno}. Finally, we conclude in Section~\ref{sec:conclusion}.

\section{The model}\label{sec:model}

The hidden sector is gauged under a dark $SU(2)_X$, and $X_\mu^a=(X_\mu^1,X_\mu^2,X_\mu^3)$ is the corresponding gauge field. We introduce the real scalar $\phi_a=(G_1,G_2,s)^T$, which is a triplet under $SU(2)_X$ and a singlet under the SM gauge group $SU(2)_L\times U(1)_Y$. The Lagrangian is written as
\be\label{Ltree}
\mL\supset-\frac14X_{\mu\nu}^aX_a^{\mu\nu}+\frac12D_\mu\phi^aD^\mu\phi^a-V_0(\phi,H)\, ,
\ee
where the dark covariant derivative is 
\be
(D_\mu\phi)_a=\partial_\mu\phi_a+g_X\epsilon^{abc}X_\mu^b\phi_c\, ,
\ee
with $g_X$ being the dark gauge coupling, and the field strength tensor is
\be
X_{\mu\nu}^a=\partial_\mu X_\nu^a-\partial_\nu X_\mu^a+g_X\epsilon^{abc}X_\mu^bX_\nu^c\, ,
\ee
with $\epsilon^{abc}$ being the 3-dimensional Levi-Civita tensor under the convention $\epsilon^{123}=1$.

The tree-level joint potential for $\phi$ and the SM Higgs doublet $H=\left(G^+,(h+iG^0)/\sqrt{2}\right)^T$ reads
\be
V(H,\phi)=\lambda_h|H|^4+\frac{\lambda_s}{4}\left(\phi^T\phi\right)^2+\frac{\lambda_{hs}}{2}|H|^2\left(\phi^T\phi\right)\, ,
\ee
which contains only dimensionless parameters. Consequently, all symmetries, including conformal invariance, $SU(2)_X$, and $SU(2)_L\times U(1)_Y$, are preserved at this order. However, radiative corrections change this picture. Working in the unitary gauge, $H\to (0,h/\sqrt{2})^T$ and $\phi\to(0,0,s)^T$, and assuming the $SU(2)_X$ scale $\ave{s}=w$ much higher than the EW scale $\ave{h}=v=246$ GeV, we can adopt the sequential symmetry breaking approximation~\cite{Chataignier:2018kay}. In this limit, the one-loop potential is
\be\label{CC_V}
V_1(h,s)=\frac{3g_X^4}{16\pi^2}s^4\left(\log\frac{s}{w_0}-\frac14\right)+\frac{\lambda_{hs}}{4}h^2s^2+\frac{\lambda_h}{4}h^4\, .
\ee
The logarithmic potential along the $s$ direction is generated by $X_\mu^a$ via the CW mechanism, yielding the vacuum expectation value (VEV) $w\approx w_0$. This breaks the conformal symmetry, and also breaks $SU(2)_X$ down to $U(1)_X$, giving a mass $m_X=g_Xw$ to $X_\mu/X_\mu^\dagger=(X_\mu^1\pm iX_\mu^2)/\sqrt{2}$, while $X_\mu^3$ remains massless, denoted as the dark photon $A'_\mu$. Note $X_\mu/X_\mu^\dagger$ is charged $\mp1$ under $U(1)_X$ but neutral under the SM gauge group. The physical scalar $s$ also acquires a mass $m_s\approx\sqrt{3}g_X^2w/(2\pi)$. Following the $SU(2)_X$ breaking, the negative portal coupling $\lambda_{hs}\approx-m_h^2/w^2$ induces a negative mass-squared term $-m_h^2h^2/4$ for the Higgs field, triggering the EW symmetry breaking.

The portal coupling $\lambda_{hs}$ also causes the mixing between the physical scalar bosons $s$ and $h$, described by
\be
\begin{pmatrix}h\\ s\end{pmatrix}\xrightarrow[\rm eigenstates]{\rm to~mass} U\begin{pmatrix}h\\ s\end{pmatrix}\, ,\quad U=\begin{pmatrix}\cos\theta&\sin\theta\\ -\sin\theta&\cos\theta\end{pmatrix}\, ,
\ee
with $|\theta|<\pi/4$ being the mixing angle that is determined by the potential in \Eq{CC_V}~\cite{Liu:2024fly}. Given the measured values of Higgs VEV $v$ and mass $m_h$, our model possesses only two free parameters, which we choose as the mass eigenvalues $m_X$ and $m_s$ that are closely related to DM phenomenology. All other parameters, such as $g_X$, $w$ (or $w_0$), and $\theta$, can then be derived analytically. Following the procedure outlined in Ref.~\cite{Liu:2024fly}, we obtain
\be
w=\frac{1}{2 \sqrt{2} \pi  m_h m_s}\sqrt{\mD\pm\sqrt{\mD^2-36 m_h^2 m_s^2m_X^8}}\, .
\ee
for $m_s\lessgtr m_h$, where $\mD=3 m_X^4 \left(m_h^2+m_s^2\right)-4\pi ^2 m_h^2 m_s^2 v^2$. A physical solution for $w$ exists only if
\be\label{solution}
m_X > \sqrt{\frac{2\pi m_h m_sv}{\sqrt{3}|m_s-m_h|}}\, ,
\ee
which represents an implicit consistency condition in the $(m_X,m_s)$ parameter space. Once $w$ is determined, $g_X=m_X/w$ is also derived, and $\theta$ can be resolved from the relation
\be
w=\frac{m_h^2\cot\theta+m_s^2\tan\theta}{|m_h^2-m_s^2|}v\, .
\ee
This simplifies to $\tan\theta\approx(v/w)m_h^2/(m_h^2-m_s^2)$ for $w\gg v$~\cite{Iso:2009nw}. Further details of this parametrization are provided in Appendix~\ref{app:parametrization}.

After symmetry breaking, expanding the Lagrangian in \Eq{Ltree} yields the interactions among the physical particles. The triple couplings are
\be
\mL_3\supset\frac{2m_X^2}{w}sX_\mu^\dagger X^\mu
+ig_X(X_{\mu\nu}^\dagger X^\mu-X_{\mu\nu}X^{\dagger\mu})A'^{\nu}
+ig_X\partial^\mu A'^\nu(X_\mu^\dagger X_\nu-X_\mu X_\nu^\dagger)\, ,
\ee
and the quartic couplings are
\begin{multline}
\mL_4\supset g_X^2s^2X_\mu^\dagger X^\mu+g_X^2\left(X^{\dagger\mu}X^\nu A'_\mu A'_\nu-X_\mu^\dagger X^\mu A'_\nu A'^\nu\right)\\
+\frac{g_X^2}{2}\left(X_\mu^\dagger X_\nu X^{\dagger\mu}X^\nu-X_\mu^\dagger X^\mu X_\nu^\dagger X^\nu\right)\,.
\end{multline}
The scalar boson $s$ also couples to SM fermions and gauge bosons via the mixing angle $\theta$, and conversely, the Higgs boson $h$ couples to $X$ via the same mixing. The vertices between $h$ and $s$ are less relevant to DM dynamics and hence listed in Appendix~\ref{app:parametrization}.

At tree-level, $s$ does not couple to the dark photon because it is neutral under $U(1)_X$; however, loop corrections involving the charged $X$-bosons generate the dimension-five operator \be
\mL_{sA'A'}\approx\frac{7g_X^3}{32\pi^2m_X}sF'_{\mu\nu}F^{\prime\mu\nu}\, ,
\ee
with $F'_{\mu\nu}\equiv\partial_\mu A'_\nu-\partial_\nu A'_\mu$, thereby inducing the $sA'A'$ coupling. This is analogous to the $W$-loop-induced Higgs-photon coupling $hF_{\mu\nu}F^{\mu\nu}$ in the SM.

\section{Thermal history}\label{sec:history}

In the early Universe, the plasma induces a temperature dependence in the scalar potential. This usually leads to symmetry restoration at high temperatures and symmetry breaking at lower temperatures, thus driving a cosmic phase transition~\cite{Dolan:1973qd}. A distinguishing feature of CC theories is that they generally result in FOPTs.

\subsection{Classification of the evolution patterns}

Because of the smallness of $|\lambda_{hs}|$, the thermal corrections of the dark and SM sectors can be treated separately. The finite-temperature effective potential is therefore
\be\label{VT}
V_T(h,s,T)=V_1(h,s)+V_X^T(s,T)+V_h^T(h,T)\,,
\ee
where each contribution is detailed below. The thermal contribution from the dark sector,
\be
V_X^T(s,T)=\frac{3T^4}{\pi^2}J_B\left(\frac{g_X^2s^2}{T^2}\right)
-\frac{g_X^3}{6\pi}T\left[\left(s^2+T^2\right)^{3/2}-s^3\right]\, ,
\ee
includes the one-loop thermal integral together with the daisy-resummation correction for the longitudinal modes of the $SU(2)_X$ gauge bosons. For the SM sector, the thermal potential is
\be
V_h^T(h,T)=\sum\frac{n_iT^4}{2\pi^2}J_{B/F}\left(\frac{M_i^2(h)}{T^2}\right)+V_{\rm SM}^{\rm ds}(h,T),
\ee
where the sum runs over the SM degrees of freedom (d.o.f.). Dominant contributions come from the EW gauge bosons, the top quark, and the scalars:
\be
\begin{dcases}
~M_W^2(h)=g^2h^2/4\,,&n_W=2\times3=6\, ;\\
~M_Z^2(h)=(g^2+g'^2)h^2/4\,,&n_Z=3\, ;\\
~M_t^2(h)=y_t^2h^2/2\,,&n_t=N_c\times 4=12\, ;\\
~M_h^2(h)=3\lambda_hh^2\,,&n_h=1\, ;\\
~M_G^2(h)=\lambda_hh^2\,,&n_G=3\, ,\\
\end{dcases}
\ee
where $g$ and $g'$ are the gauge couplings of the $SU(2)_L$ and $U(1)_Y$ group, respectively. The daisy resummation term $V_{\rm SM}^{\rm ds}(h,T)$ can be found from Ref.~\cite{Carrington:1991hz}. The thermal integrals are
\be
J_{B/F}(y)=\pm\int_0^\infty\d xx^2\log\left(1\mp e^{-\sqrt{x^2+y}}\right)\, ,
\ee
where the upper (lower) sign corresponds to bosons (fermions).

To illustrate the early Universe history, here we expand the thermal potential near the origin $(h,s)=(0,0)$, which yields
\be\label{VT_approx}
\text{\Eq{VT}}\approx V_1(h,s)+\frac{g_X^2T^2}{4}\left(1-\frac{g_X}{\pi}\right)s^2+\frac{c_hT^2}{2}h^2\, ,
\ee
where $c_h=(3g^2+g'^2)/16+y_t^2/4+\lambda_h/2\approx0.4$. The CC potential $V_1(h,s)$ is flat at the origin: both the first and second order derivatives vanish. Consequently, the $T^2$-terms always form a local minimum at $(0,0)$, in which the Universe stays when $T\gg w$. As $T$ drops to the critical temperature $T_c$ (which is typically $\lesssim w$), a global minimum (the true vacuum) exists at $h$, $s\neq0$, which eventually aligns with the zero-temperature VEV $(v,w)$. However, as the origin remains a local minimum, a smooth transition to the true vacuum is always forbidden. Instead, the Universe has to tunnel discontinuously to the true vacuum, yielding a FOPT.

The cosmic evolution is governed primarily by the FOPT dynamics of the high-scale sector~\cite{Iso:2017uuu}. We therefore focus on the $O(3)$-symmetric bounce solution action $S_3/T$ derived from the $s$-direction thermal potential
\be\label{VTs}
V_T(s,T)=V_1(0,s)+V_X^T(s,T)\, ,
\ee
which determines the vacuum tunneling rate from $s=0$ to $s\neq0$ per unit volume as~\cite{Linde:1981zj}
\be\label{eq:tunnelingrate}
\Gamma(T)\sim T^4\left(\frac{S_3}{2\pi T}\right)^{3/2}e^{-S_3/T}\, . 
\ee
This leads to the nucleation of bubbles containing the true vacuum, resulting in a decreasing volume fraction $F(T)=e^{-I(T)}$ of the false vacuum as $T$ decreases, where 
\be\label{eq:I(T)}
I(T)=\frac{4\pi}{3}\int_T^{T_c}\frac{\Gamma(T')\d T'}{T' H(T')}\left( R_0(T') + \frac{1}{T'} \int_T^{T'}\frac{v_w\d T''}{H(T'')}\right)^3\, ,
\ee
where $H(T)$ is the Hubble constant determined by the first Friedmann equation
\be
H(T)=\sqrt{\frac{8\pi}{3M_{\rm Pl}^2}\left(\Delta V_T+\frac{\pi^2}{30}g_*T^4\right)}\,,
\ee
with $g_*$ the number of effective d.o.f. for energy, $v_w$ is the bubble wall expansion velocity, $\Delta V_T$ is the positive potential‑energy difference between the true and false vacua, and $R_0(T)$ the initial radius of bubble~\cite{Ellis:2019oqb}. In a supercooled FOPT, the released vacuum energy can accelerate the bubble wall to a velocity extremely close to the speed of light~\cite{Bodeker:2017cim}, therefore we take $v_w \approx 1$ as a benchmark in our FOPT analysis.

When $F(T)$ drops to 0.71, the true vacuum bubbles are able to form an infinite connected cluster, known as percolation~\cite{rintoul1997precise}, and the corresponding temperature $T_*$ is adopted as the FOPT temperature. For a small $g_X$, the action scales as $S_3/T\propto g_X^{-3}$~\cite{Iso:2017uuu}, resulting in a highly suppressed tunneling rate and hence a supercooled transition with $T_* \ll w$. In extreme cases, if percolation cannot occur before $T_{\rm QCD}\approx85$ MeV, the QCD phase transition takes place first, dramatically altering the structure of \Eq{VTs}. The resulting cosmological histories fall into two main patterns, comprising four distinct types classified in Ref.~\cite{Liu:2024fly}.

If $F(T)$ reaches 0.71 at $T_*>T_{\rm QCD}$, the evolution follows the {\bf normal pattern}. The high-scale $SU(2)_X$ and conformal FOPT occurs first at $T_*$, and then affects the EW phase transition. This pattern can be further classified depending on the relation between $T_*$ and the EW characteristic temperature $T_{\rm ew} \equiv m_h/\sqrt{2c_h} \approx 140$ GeV:
\begin{enumerate}
\item \underline{Type‑N1} ($T_* > T_{\rm ew}$), $SU(2)_X$ breaks via a FOPT at $T_*$, while the EW symmetry breaks via a smooth crossover near $T_{\rm ew}$;
\item \underline{Type‑N2} ($T_* < T_{\rm ew}$), a joint $SU(2)_X$-EW FOPT takes place at $T_*$.
\end{enumerate}
If $\Gamma(T)$ is sufficiently suppressed such that $F(T) > 0.71$ persists down to $T_{\rm QCD}$, the evolution enters the {\bf inverted pattern}. The Universe stays in the symmetric phase $(0,0)$ down to $T_{\rm QCD}$, and a first-order QCD confinement transition is triggered by six-flavor massless quarks~\cite{Pisarski:1983ms,Braun:2006jd,Guan:2024ccw}. This induces a Higgs VEV $h = v_{\rm QCD} \approx 100$ MeV via the top‑quark condensate~\cite{Witten:1980ez,Iso:2017uuu}, reshaping the potential along the $s$-direction via
\be
\delta V_{\rm QCD}(s,T)=\frac{\lambda_{hs}}{4}h^2s^2\Big|_{h\to v_{\rm QCD}}\approx-\frac{m_h^2v_{\rm QCD}^2}{4w^2}s^2\, ,
\ee
adding to the thermal potential $V_T(s,T)$ in \Eq{VTs}. This negative mass-squared term for $s$ lowers or even dismisses the barrier formed by the $T^2s^2$ term, thereby qualitatively altering the tunneling behavior. The subsequent dynamics is determined by the ordering between $T_{\rm QCD}$ and the rolling temperature $T_{\rm roll} \equiv m_h v_{\rm QCD} / (g_X w\sqrt{1-g_X/\pi})$ that the barrier disappears:
\begin{enumerate}
\item \underline{Type‑I1} ($T_{\rm QCD} > T_{\rm roll}$), after the QCD FOPT, an $SU(2)_X$ FOPT occurs at $T_* \approx T_{\rm roll}$, which then drives the Higgs field from $v_{\rm QCD}$ to its final VEV $v$;\footnote{Lattice simulations show that the transition still proceeds as a FOPT even when $T$ is very close to $T_{\rm roll}$ and the barrier is sufficiently low~\cite{Dutka:2025oqt}.}
\item \underline{Type‑I2} ($T_{\rm QCD} < T_{\rm roll}$), the QCD FOPT instantaneously triggers a joint $SU(2)_X$-EW FOPT at $T_* = T_{\rm QCD}$, and modeling of QCD phase transition dynamics is necessary~\cite{Sagunski:2023ynd}.
\end{enumerate}

\subsection{Transition regimes in the $(m_X,m_s)$ plane}

Using the complete one‑loop thermal potential \Eq{VT} and solving the bounce action $S_3$ numerically, we calculate the FOPT dynamics within the CC $SU(2)_X$ model and classify each point in the $(m_X,m_s)$ parameter space according to its cosmological evolution type. All four types of evolution history involve an $SU(2)_X$ FOPT. We characterize this transition by two standard parameters: the ratio of latent heat to the radiation energy density
\be
\alpha=\frac{1}{\pi^2g_*T_*^4/30}\left.\left(\Delta V_T-\frac{T}{4}\frac{\d\Delta V_T}{\d T}\right) \right|_{T_*}\, ;
\ee
and the Hubble time relative to the duration of the transition,
\be
\frac{\beta}{H_*} = T \left.\frac{\d (S_3/T)}{\d T} \right|_{T_*}\, ,
\ee
where $H_*=H(T_*)$.

For a supercooled transition, it is convenient to express the vacuum energy density in terms of a characteristic temperature scale, $T_\Lambda$, defined by $\pi^2 g_* T_\Lambda^4/30 = \Delta V_T \big|_{T=0}$. When $T_*\ll T_\Lambda$, the Universe enters a vacuum-dominated era before the FOPT, i.e., a period of thermal inflation. The subsequent transition releases this vacuum energy and reheats the Universe from $T_*$ to $T_{\rm rh}$~\cite{Hambye:2018qjv},
\be
T_{\mathrm{rh}}= T_\Lambda\times \min \left(1,~ \frac{\Gamma_{\rm eff}}{H\left(T_*\right)} \right)^{1/2},
\ee
where $\Gamma_{\rm eff}=\Gamma_h \sin ^2 \theta+\Gamma_s \cos ^2 \theta$ with $\Gamma_h$ and $\Gamma_s$ being the decay widths of the $h$ and $s$ bosons, respectively. In the supercooled regime ($\alpha \gg 1$), $T_\Lambda\approx (1+\alpha)^{1/4}T_*$. Another relevant parameter is $(8\pi)^{1/3}/(R_*H_*)$, where $R_*=n_b^{-1/3}$, with the bubble number density
\be
n_b(T)=T^3\int_T^{T_c}\d T'\frac{\Gamma(T')}{T'^4H(T')}F(T')\,.
\ee
For a prompt FOPT, $(8\pi)^{1/3}/(R_*H_*)$ is well approximated by the usual $\beta/H_*$ parameter~\cite{Enqvist:1991xw}, whereas in a supercooled transition the two can differ significantly.

\begin{figure*}
\centering
\includegraphics[width=0.95\linewidth]{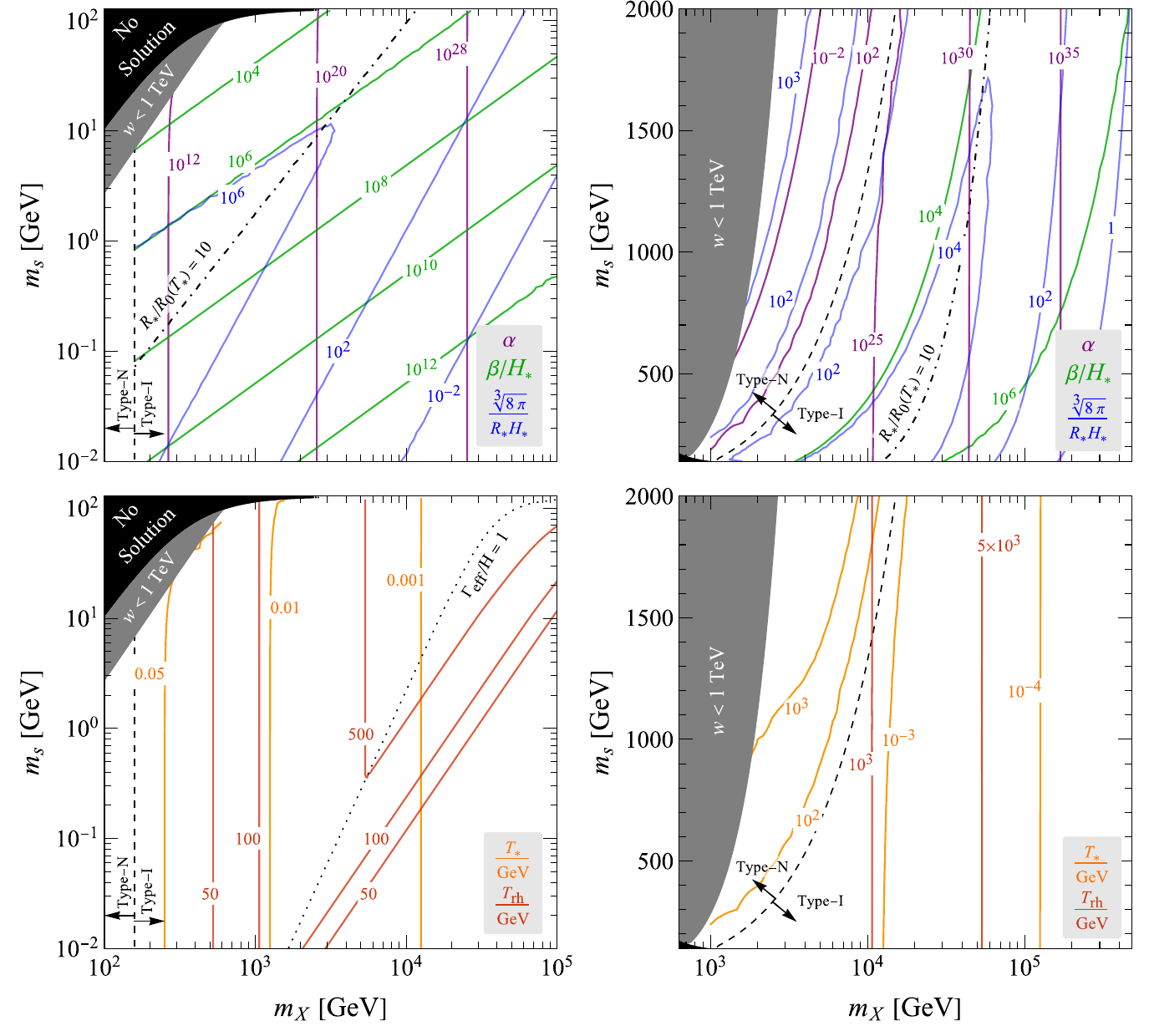}
\caption{Contours of FOPT characteristic parameters $\alpha$ (purple), $\beta /H$ (green), $(8\pi)^{1/3}/(R_*H_*)$ (blue), $T_*$ (orange), and $T_{\text{rh}}$ (red) for $m_s<m_h$ (left) and $m_s>m_h$ (right). The black and gray shaded areas denote the unphysical and $w < 1 \text{ TeV}$ regions. The black dashed lines are the boundary of different thermal history patterns. Black dot-dashed lines denote $R_*/R_0(T_*)=10$ in the top panel, while dotted line denotes $\Gamma_{\rm eff}/H=1$ in the bottom panel.}
\label{fig:foptparam}
\end{figure*}

The FOPT parameters are shown in Fig.~\ref{fig:foptparam}, where the left and right panels correspond to $m_s<m_h$ and $m_s>m_h$, respectively. The top panel displays the contours of $\alpha$ (purple), $\beta/H_*$ (green), and $(8\pi)^{1/3}/(R_*H_*)$ (blue), while the bottom panel displays the characteristic temperatures: the percolation temperature $T_*$ (orange) and the reheating temperature $T_{\rm rh}$ (red). In both panels, the black dashed line marks the boundary between the Type-I and Type-N evolution history patterns. The gray shaded regions indicate $w<1$ TeV, where the sequential symmetry breaking assumption is invalid. The black shaded regions correspond to parameters for which \Eq{solution} has no solution, and hence no physical $w$ exists. Black dot-dashed lines in the top panel denote $R_*/R_0(T_*)=10$, and the regions below the line correspond to smaller values. Black dotted line in the bottom panel denotes $\Gamma_{\rm eff}/H=1$.

Most of the parameter space in the left panel of Fig.~\ref{fig:foptparam} corresponds to Type-I1 evolution. This scenario features an ultra-supercooled FOPT with $T_* = \mO(1-100)$ MeV, $\alpha = \mO(10^{12}-10^{28})$, and an ultrashort duration $\beta/H_* = \mO(10^4-10^{12})$. In the Type-I region where $R_*/R_0(T_*) < 10$, $(8\pi)^{1/3}/(R_*H_*)$ differs significantly from $\beta/H_*$ when $R_0$ becomes important in \Eq{eq:I(T)}, whereas in Type-N the two quantities coincide. The contours of $\alpha$ and $T_*$ are nearly independent of $m_s$ (almost parallel to the $m_s$ axis), since both the vacuum energy difference and the FOPT temperature $T_*$ ($\approx T_{\rm roll}$ in Type-I1) are dominated by $m_X = g_X w$. The kink in the reheating temperature contour of $T_{\rm rh}=500$ GeV marks the transition to the slow-reheating regime, where $\Gamma_{\rm eff} < H(T_*)$. The $\beta/H$ contours closely follow those of the gauge coupling $g_X$; for instance, $\beta/H = 10^4$ ($10^8$) roughly overlaps with $g_X = 10^{-1}$ ($10^{-3}$). A detailed explanation is provided in Appendix~\ref{app:PTDyna}.

In the right panel of Fig.~\ref{fig:foptparam}, both Type-N and Type-I1 evolutions appear, separated by the black dashed line. In the Type-N region, the FOPT occurs at $T_* = \mO(10^2 - 10^3)$ GeV with mild supercooling, and the $\beta/H_*$ and $(8\pi)^{1/3}/(R_*H_*)$ values coincide. To the right of the dashed line, the Type-I1 pattern yields an ultra-supercooled FOPT, with $T_*$ dropping to $\mO(0.1-100)$ MeV and $\beta/H$ reaching $\mO(10^4-10^6)$. In this domain, $\alpha$ and $T_*$ are set by $m_X$, so their contours remain nearly parallel to the $m_s$ axis. 

It should be emphasized that the analysis of the FOPT parameters across both panels of Fig.~\ref{fig:foptparam} relies on the one-loop effective potential and the semiclassical tunneling rate; a more rigorous calculation incorporating higher-order corrections to the effective potential~\cite{Christiansen:2025xhv,Liu:2026ask,Qin:2024dfp} and the functional determinant of the tunneling rate~\cite{Ai:2023yce, Ekstedt:2023sqc, Gould:2024chm,Wang:2026ifp} would yield more accurate values for $T_*$, $\alpha$, and $\beta/H$. Nevertheless, as we will show, these corrections have only a minor impact on the DM dynamics. This is because WIMPs are in thermal equilibrium and hence insensitive to the FOPT history, while for supercooled DM, which lies in the small-coupling regime (e.g., $g_X=\mO(10^{-2}-10^{-3})$), the two-loop contributions remain small~\cite{Christiansen:2025xhv}. In addition, since $\beta/H\gg 1$ throughout the parameter space of interest, $S_3/T$ is highly sensitive to temperature, implying that the decay rate is dominated by the exponential factor and that the semiclassical approximation is well justified. We therefore expect our results to be robust.

\section{Dark matter scenarios}\label{sec:dm}

Based on the thermal history outlined above, the CC $SU(2)_X$ model gives rise to two viable DM scenarios for the $X^\pm$ gauge bosons: WIMP freeze-out and supercooled DM, with the production mechanism determined by the thermalization history following the FOPT. We examine each of these possibilities in turn. The model's triplet structure also admits a monopole candidate; however, we find that it cannot account for the observed relic abundance, as shown in Appendix~\ref{app:monopole}.

\subsection{Resolving the relic abundance}

Before the $SU(2)_X$ FOPT, the gauge bosons $X^\pm$ and $A'$ are massless, with an abundant number density $\sim T^3$ in the thermal bath. The equilibrium yield of $X^\pm$ in the symmetric phase can be obtained from the massless Bose-Einstein distribution, i.e.,
\be\label{Yeqs}
Y_{\rm eq}^{\mathfrak{s}}\equiv\frac{3\zeta_3T^3/\pi^2}{2\pi^2 g_{*s} T^3/45}=\frac{135\zeta_3}{2\pi^4g_{*s}}\, ,
\ee
where $\zeta_3 \approx 1.202$, $g_{*s}$ is the number of effective d.o.f. for entropy. Note that only one sign of $X^\pm$ ($X^+$ or $X^-$) is counted in \Eq{Yeqs}. The FOPT drastically alters this picture through two effects: mass gain and entropy injection.

During the $SU(2)_X$ FOPT, the charged bosons $X^\pm$ acquire a mass $m_X^* = g_X w_*$ inside the bubble, where $w_*$ is the true vacuum value of $s$ at $T_*$. In the supercooled regime of interest, $T_*\ll w$, so
$w_*\approx w$ and hence $m_X^*\approx m_X$. In a moderate FOPT, if $m_X^* \gg T_*$, most $X^\pm$ bosons lack sufficient kinetic energy to penetrate the bubble wall due to the repulsive force induced by the mass gap~\cite{Baker:2019ndr,Chway:2019kft,Hong:2020est}. However, in a supercooled FOPT, the wall velocity $v_w \approx 1$, yielding a huge Lorentz factor $\gamma_w = (1-v_w^2)^{-1/2}\gg1$. In the wall frame, the incident particles are boosted to energies $\sim \gamma_w T_* \gg m_X^*$, enabling all of them to enter the bubble~\cite{Baldes:2021vyz,Huang:2022vkf,Chun:2023ezg}. Despite complete penetration, the final abundance of $X^\pm$ is diluted by the entropy release during the post-FOPT reheating. The resulting abundance of $X^+$ or $X^-$ particles right after reheating is~\cite{Hambye:2018qjv}
\be
Y_{\rm rh}=Y_{\rm eq}^{\mathfrak{s}}\left(\frac{T_{\rm rh}}{T_\Lambda}\right)\left(\frac{T_*}{T_\Lambda}\right)^3\, ,
\ee
which serves as the initial condition for subsequent DM evolution. For the supercooled transition we consider, $Y_{\rm rh}$ is typically highly suppressed and can be neglected.

Following the FOPT, the scalar fields undergo coherent oscillations around the vacuum and decay into SM light particles, such as $e^+e^-$, $\nu\bar\nu$, or $b\bar b$, generating a thermal plasma, and the Universe then enters a radiation era. In the parameter space of interest, $s$ can thermalize via the interactions with the SM particles. The $X^\pm$ bosons can then be produced through annihilations of $s$ or SM particles. For a prompt reheating, $\Gamma_{\rm eff}>H_*$, the evolution of $X^\pm$ is then described by the Boltzmann equation
\be\label{YX}
\frac{\d Y}{\d z}=\sqrt{\frac{\pi g_{*s}^2}{45g_*}}\frac{M_{\rm Pl}m_X}{z^2}\ave{\sigma v_{\rm rel}}\left[(Y_{\rm eq}^{\mathfrak{b}})^2-Y^2\right]\, ,
\ee
where $M_{\rm Pl}=1.22\times10^{19}$ GeV is the Planck scale, $Y$ denotes the $X^+$ or $X^-$ yield (i.e., number density over the entropy density) after the FOPT,
\be
Y_{\rm eq}^{\mathfrak{b}}\equiv\frac{3m_X^2TK_2(m_X/T)/(2\pi^2)}{2\pi^2 g_{*s} T^3/45}=\frac{135z^2K_2(z)}{4\pi^4g_{*s}}\, 
\ee
is the equilibrium yield in the $SU(2)_X$-breaking phase with $K_2(z)$ the modified Bessel function of the second kind, $z\equiv m_X/T$, and $\ave{\sigma v_{\rm rel}}$ is the thermally averaged annihilation cross section whose details are given in Appendix~\ref{app:xsection}. The dominant annihilation channels contributing to $\ave{\sigma v_{\rm rel}}$ are
\be\label{ann}
X^+X^-\to A'A' / ss;\quad X^+X^-\to W^+W^-/ZZ/t\bar t\, ,
\ee
where annihilations into SM final states proceed via off-shell $h$ or $s$ bosons.

The evolution described by \Eq{YX} begins at $z_{\rm rh} \equiv m_X/T_{\rm rh}$ with the initial condition $Y(z_{\rm rh})=Y_{\rm rh}$. The resulting relic abundance is determined by the relationship between the $X^\pm$ annihilation rate
\be
\Gamma_{\rm ann}\equiv \left.\left(\frac{2\pi^2}{45} g_{*s} T_{\rm rh}^3\right)Y_{\rm eq}^{\mathfrak{b}}\ave{\sigma v_{\rm rel}}\right|_{T_{\rm rh}}
\ee
and the Hubble parameter $H_{\rm rh}\equiv H(T_{\rm rh})$ at the reheating temperature. If $\Gamma_{\rm ann}>H_{\rm rh}$, the $X^\pm$ bosons rapidly thermalize and subsequently undergo conventional thermal freeze-out. This yields the familiar WIMP miracle~\cite{Bertone:2004pz}
\be\label{WIMP_sim}
\Omega_{\rm fo}h^2\approx2\times\frac{2.55\times10^{-10}~{\rm GeV}^{-2}}{\ave{\sigma v_{\rm rel}}}\, .
\ee

Conversely, if $\Gamma_{\rm ann}<H_{\rm rh}$, the $X^\pm$ bosons never reach thermal equilibrium. Their abundance is instead set by a freeze-in mechanism, with the asymptotic yield given by~\cite{Wong:2023qon}
\be
Y_\infty\approx Y_{\rm rh}+\frac{135\sqrt{5}M_{\rm Pl}m_X\ave{\sigma v_{\rm rel}}}{128\pi^{13/2}g_{*s}\sqrt{g_*}}(1+2z_{\rm rh})e^{-2z_{\rm rh}}\, ,
\ee
and the corresponding relic density is
\be
\Omega_{\rm sc}h^2=2Y_\infty s_0 m_X\frac{8\pi}{3M_{\rm Pl}^2}\left(\frac{h}{H_0}\right)^2\, ,
\ee
where $s_0 \approx 2891~{\rm cm}^{-3}$ is the present entropy density~\cite{ParticleDataGroup:2024cfk} and $H_0/h=100~{\rm km/(s\cdot Mpc)}$. This production mechanism is referred to as supercooled DM~\cite{Hambye:2018qjv,Baldes:2018emh,Wong:2023qon,Ayazi:2025sfs}.\footnote{Similar mechanism for a low inflationary reheating temperature is also possible~\cite{Lebedev:2024mbj,Arcadi:2024obp}.} In this scenario, typically $z_{\rm rh}\gtrsim20$, and the expression simplifies to the useful approximate form\footnote{Note that in the supercooled DM regime, the preexisting $A'$ population is diluted after the phase transition and does not thermalize afterward. Consequently, we include $X^+X^- \to A'A'$ as a depletion channel, while neglecting the inverse process $A'A' \to X^+X^-$.}
\be\label{sc_sim}
\Omega_{\rm sc}h^2\approx \left(\frac{m_X}{\rm TeV}\right)^2\left(\frac{z_{\rm rh}}{23}\right)\frac{e^{-2(z_{\rm rh}-23)}\ave{\sigma v_{\rm rel}}}{2.43\times10^{-10}~{\rm GeV}^{-2}}\, .
\ee
This highlights the key features that distinguish supercooled DM from WIMP freeze‑out: the relic density is proportional to $\ave{\sigma v_{\rm rel}}$, and it exhibits an exponential Boltzmann suppression from $m_X/T_{\rm rh}$. It should be noted that Eqs.~(\ref{WIMP_sim}) and~(\ref{sc_sim}) are provided for illustration; in our analysis, the relic abundance is obtained by numerically solving the full Boltzmann equation in \Eq{YX}.

We comment on the possible presence of a dark radiation background today. After the FOPT, the yield of $A'$ is diluted to be negligible by the entropy production, and $A'$ cannot re-thermalize. This is because the plasma after reheating only contains SM particles and the $s$ bosons: the SM particles do not directly couple to $A'$, while the loop-induced $sA'A'$ coupling is suppressed by the large $SU(2)_X$ breaking scale and not sufficient to thermalize $A'$. Consequently, the evolution of $A'$ is tied to that of the $X^\pm$ bosons via the unsuppressed $X^+X^-A'$ vertex. In the WIMP freeze‑out scenario, $X^\pm$ thermalizes and reproduces the dark photon bath via $X^+X^- \to A'A'$. By contrast, in the supercooled DM scenario, $X^\pm$ never reaches equilibrium, and no significant dark photon bath is regenerated after the transition.


\subsection{Viable parameter space for dark matter}

\begin{figure*}[t]
\centering
\includegraphics[width=0.49\linewidth]{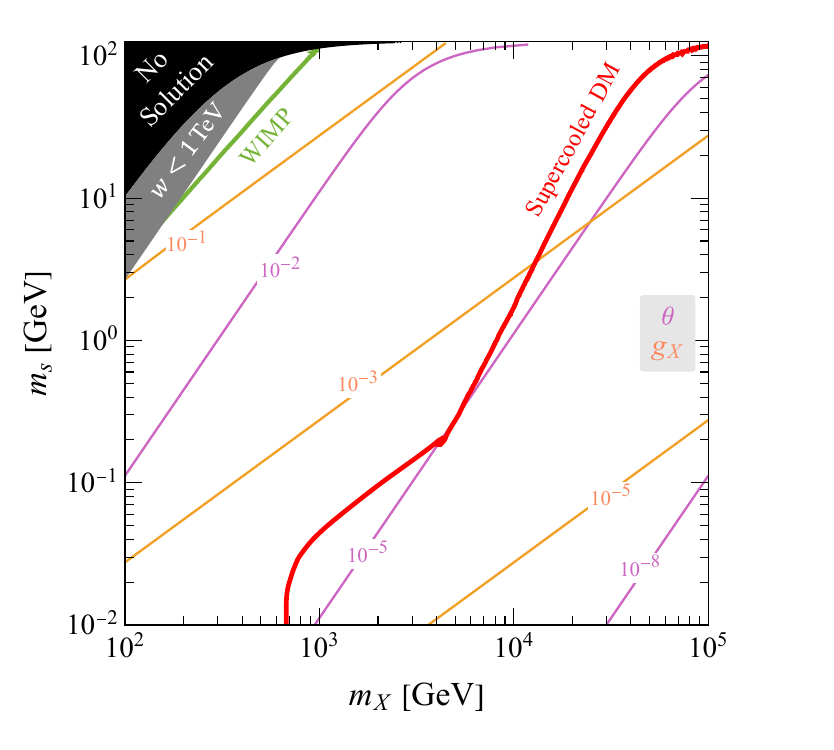}
\includegraphics[width=0.49\linewidth]{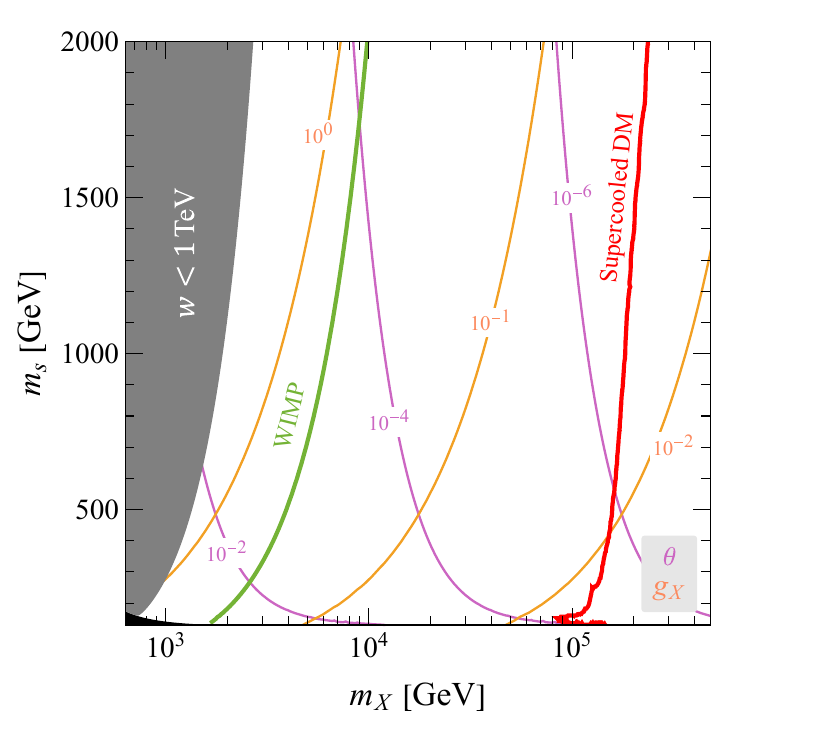}
\caption{Valid DM scenarios in the $(m_X, m_s)$ plane for $m_s < m_h$ (left) and $m_s > m_h$ (right). Bold colored curves indicate parameter space with the correct DM relic abundance: WIMP (green) and supercooled DM (red). Contours of $g_X$ and $\theta$ are shown as orange and magenta contours, respectively. All other conventions follow Fig.~\ref{fig:foptparam}.}
\label{fig:DM}
\end{figure*}

The DM parameter space in the $(m_X, m_s)$ plane is shown in Fig.~\ref{fig:DM}, where curves corresponding to the correct relic abundance are highlighted for two distinct scenarios: WIMP (green) and supercooled DM (red). In the WIMP regime, the post-FOPT reheating temperature satisfies $z_{\rm rh} \lesssim 20$, such that the $X^\pm$ bosons efficiently thermalize with the plasma, and largely erase the memory of additional nonthermal production, such as the relativistic-bubble-wall mechanism discussed in Ref.~\cite{Azatov:2021ifm}. Their final abundance is then determined by the standard freeze-out condition, leading to the familiar inverse dependence on the annihilation cross section. This region is typically associated with EW-size gauge couplings, $g_X \sim 10^{-1}$, and a DM mass $m_X$ around the TeV scale. 

The supercooled DM scenario differs significantly from the WIMP case. The smaller gauge couplings prevent the $X^\pm$ bosons from reentering thermal equilibrium, leading to a freeze-in production mechanism. While the small couplings drive a significant deviation from the typical $z_{\text{rh}} \sim 20$, the model characteristically yields $z_{\text{rh}} \sim 10$ within the instantaneous reheating regime; the high reheating temperature $T_{\rm rh}$ generated by the supercooling allows for a heavier DM mass spectrum, occupying a distinct region with $m_X =\mO(10 - 10^2)~{\rm TeV}$ and significantly reduced couplings $g_X =\mO(10^{-2}-10^{-3})$. The parameter space between the WIMP and supercooled DM lines marks a region in which the $X^\pm$ bosons are overproduced.

\section{Phenomenology}\label{sec:pheno}

\begin{figure*}[t]
\centering
\includegraphics[width=0.48\linewidth]{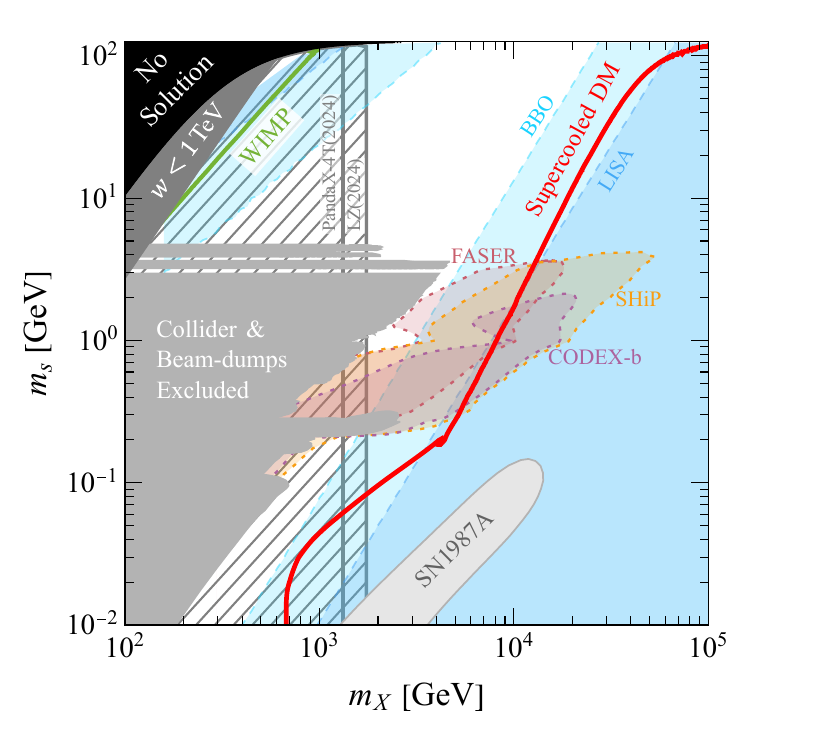}
\includegraphics[width=0.48\linewidth]{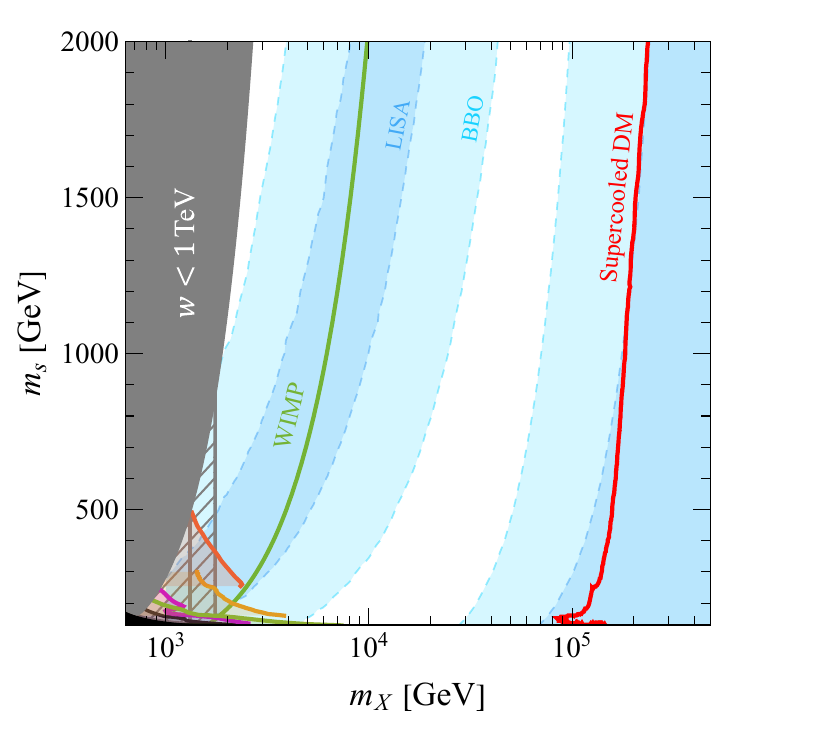}
\caption{Experimental constraints and projected sensitivities in the $(m_X, m_s)$ plane for $m_s < m_h$ (left) and $m_s > m_h$ (right). Medium gray: collider and beam-dump experiment exclusions; light gray: astrophysical bound from SN1987A; hatched gray: direct detection limits. Colored dashed contours (left) show the projected reach of LLP searches. Blue and light-blue shaded regions indicate the sensitivity of LISA and BBO to the stochastic GW background from the FOPT. Colored regions near $m_X\sim\text{TeV}$ (right) depict collider constraints and projections (see Fig.~\ref{fig:zoom} for details). Other conventions follow Figs.~\ref{fig:foptparam} and~\ref{fig:DM}.}
\label{fig:DM_ex}
\end{figure*}

We now explore the DM scenarios through their multi-messenger signatures, ranging from terrestrial particle physics experiments, including DM direct detection, LHC searches, and long-lived particle (LLP) frontiers, to cosmological gravitational wave (GW) observatories. The phenomenological viability is summarized in Fig.~\ref{fig:DM_ex}.

In the left panel of Fig.~\ref{fig:DM_ex} where $m_s < m_h$, terrestrial experiments, including LHCb~\cite{LHCb:2015nkv,LHCb:2016awg}, NA62~\cite{NA62:2020pwi,NA62:2020xlg}, CHARM~\cite{Winkler:2018qyg}, E949~\cite{BNL-E949:2009dza}, LSND~\cite{Foroughi-Abari:2020gju}, and MicroBooNE~\cite{MicroBooNE:2021sov}, constrain the mixing angle $\theta$ between $s$ and $h$, yielding the medium-gray exclusion region. Astrophysical bounds from SN1987A, based on energy-loss arguments~\cite{Winkler:2018qyg}, further restrict the parameter space. In addition, the $X^\pm$ bosons interact with each other via a long-range force mediated by the massless $A'$, potentially leaving imprints on astronomical observables such as the ellipticity of the gravitational potential of NGC720~\cite{Feng:2009mn,Agrawal:2016quu,Badziak:2022eag}. In our model, however, this constraint is subdominant, and is therefore omitted from the figure.

Another constraint arises from the DM direct detection, which sets bound on the spin-independent (SI) scattering cross section between $X^\pm$ and a nucleon $N$. This observable can be systematically analyzed within an effective field theory framework (see, e.g., Ref.~\cite{Liang:2025kkl}). In our model, the interaction is controlled by the renormalizable Higgs portal coupling, yielding~\cite{Baek:2012se}
\be
\sigma_{\text{SI}} = \frac{\mu_{XN}^2}{\pi} \frac{g_X^4 \sin^2 2\theta}{4 v^2} m_N^2 f_N^2 \left( \frac{1}{m_h^2} - \frac{1}{m_s^2} \right)^2\,,
\ee
where $\mu_{XN} = m_X m_N / (m_X+m_N)$ is the reduced mass of the DM-nucleon system, and $f_N \approx 0.3$ \cite{Hoferichter:2015dsa} is the effective nucleon coupling. As $\theta$ decreases with increasing $w$, this cross section is naturally suppressed for heavier DM. Comparing our parameter space with the latest limits from LZ~\cite{LZ:2024zvo} and PandaX-4T~\cite{PandaX:2024qfu} yields the hatched gray exclusion region shown in the figure, where the entire WIMP scenario is now ruled out.

The supercooled DM scenario remains viable under current experimental constraints and can be probed through FOPT GWs at future space-based interferometers such as LISA~\cite{LISA:2017pwj} and BBO~\cite{Crowder:2005nr}; TianQin~\cite{TianQin:2015yph} and Taiji~\cite{Ruan:2018tsw} are expected to reach comparable sensitivity. For these projections, we calculate the GW spectrum using standard fits~\cite{Caprini:2015zlo,Caprini:2019egz}, following the implementation of Ref.~\cite{Liu:2024fly} except for the characteristic length scale discussed below, and require ${\rm SNR}>50$ for an integration time of $0.75\times4$ years. While Ref.~\cite{Liu:2024fly} argued that the $m_s < m_h$ regime yields an extremely strong but short-duration FOPT (i.e., large $\beta/H_*$) and thus an undetectable GW signal, we find that its calculation omitted the initial bubble radius $R_0(T)$ in \Eq{eq:I(T)}, which becomes crucial in an ultra-supercooled FOPT.\footnote{We thank Felix Br\"{u}mmer and Giacomo Ferrante for pointing this out to us.} Here $R_0(T_*)$ is not negligible relative to $R_*$, so the usual relation $(8\pi)^{1/3}/(R_*H_*)\simeq\beta/H_*$ is no longer accurate and the GW scale is set directly by $R_*$. Including $R_0(T)$ and using $(8\pi)^{1/3}/(R_*H_*)$ therefore enhances the predicted amplitude, bringing it into the detectable range.

Further indirect detection is possible through LLP searches for the neutral scalar $s$. The supercooled DM scenario occupies the large-$w$ region, where the mixing angle $\theta$ between $s$ and $h$ is highly suppressed. This leads to a correspondingly small total decay width $\Gamma_s \propto \sin^2 \theta$, rendering $s$ sufficiently long-lived to be accessible at future LLP searches. The colored dashed contours in Fig.~\ref{fig:DM_ex} indicate the projected sensitivities of experiments such as FASER~\cite{FASER:2018eoc}, CODEX-b~\cite{Gligorov:2017nwh}, and SHiP~\cite{SHiP:2020vbd}.

\begin{figure}
\centering\includegraphics[scale=0.55]{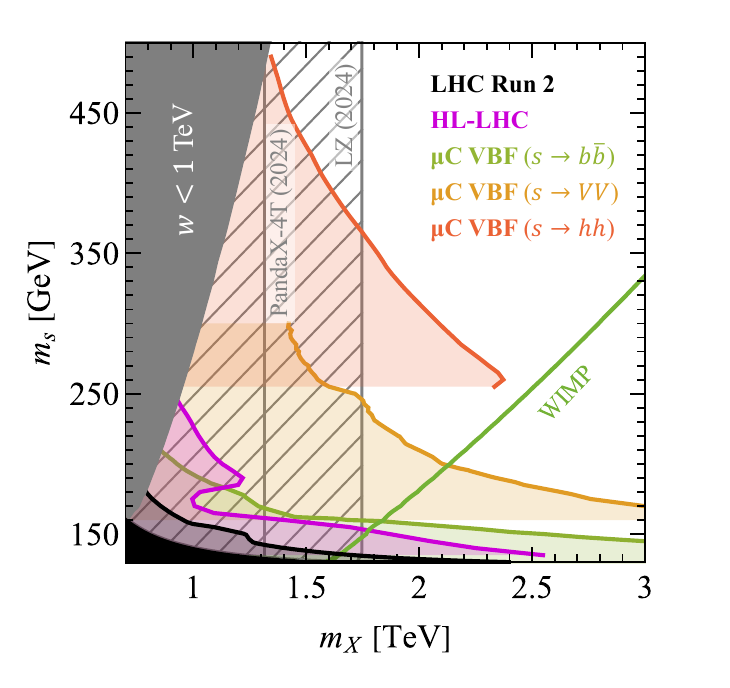}
\caption{Zoomed-in view of the right panel of Fig.~\ref{fig:DM_ex}. Shaded regions show collider constraints and projections: LHC Run 2 (semi-transparent black), HL-LHC extrapolation (purple), and a 10 TeV muon collider ($\mu$C) with $s$ produced via VBF and decaying to $b\bar{b}$ (light green), $VV$ (orange), and $hh$ (red). Conventions follow Fig.~\ref{fig:DM_ex}.}
\label{fig:zoom}
\end{figure}

In the right panel of Fig.~\ref{fig:DM_ex} ($m_s > m_h$), both the WIMP and supercooled DM scenarios are allowed by current experimental limits, although WIMP masses below $\sim 2$ TeV are excluded by direct detection. Both DM scenarios can be efficiently probed via GWs from the FOPT, as in $m_s < m_h$ scenario. Detectable signals fall within the sensitivity reach of future space-based interferometers such as LISA and BBO, as delineated by the blue and light-blue regions. Another probe for the WIMP scenario in the $m_s > m_h$ regime is the prompt decay of $s$, as shown in the right panel of Fig.~\ref{fig:DM_ex} (see Fig.~\ref{fig:zoom} for a detailed view). Current bounds are displayed as black contours, which combine LHC Run 2 constraints from Higgs signal strength measurements~\cite{ATLAS:2024fkg} ($36.1-139~\text{fb}^{-1}$) and BSM $s \to ZZ$ searches~\cite{CMS:2018amk} ($35.9~\text{fb}^{-1}$). Extrapolating these to the HL-LHC ($3000~\text{fb}^{-1}$) yields the purple contour, which offers significant coverage for $m_h \lesssim m_s \lesssim 2m_h$, with a notable dip near $m_s \sim 170$ GeV due to branching ratio suppression~\cite{Djouadi:2005gi}. Probing deeper into the parameter space requires a future 10 TeV muon collider ($\mu$C), where vector boson fusion (VBF) production of $s$ can reach sensitivities down to $\theta \sim 10^{-2}$. The reach varies across decay channels, reflecting the dominant branching fractions of $s$: $s \to b\bar{b}$ (light green) and $s \to VV$ (orange) offer significantly broader coverage than $s \to hh$ (red). For a detailed phenomenological analysis of these VBF processes at the $\mu$C, see Ref.~\cite{Liu:2024fly}.

In the WIMP scenarios with a dark-photon thermal bath after the FOPT, we also examine the $N_{\rm eff}$ constraint. We first verify that $X$ always remains in kinetic equilibrium with $A'$ through the unsuppressed scattering $X^\pm A'\to X^\pm A'$ over the relevant temperature range. We then determine the kinetic decoupling temperature from the momentum-transfer rate of $X^\pm$ elastic scattering off SM particles~\cite{Bringmann:2006mu,Visinelli:2015eka,Gondolo:2012vh}. We find that typically $m_X/T_{\rm kd}=\mO(10^3)$ in our parameter space, so kinetic decoupling occurs when $X$ is already non-relativistic and chemically frozen out. As a result, after kinetic decoupling, the dark thermal bath temperature is $T_{A'}\approx T_{\rm SM}\cdot(g_{*s,\text{SM}}(T_{\rm SM})/g_{*s,\text{SM}}(T_{\rm kd}))^{1/3}$. Along the $m_s>m_h$ WIMP line we obtain $\Delta N_{\rm eff}\simeq0.071-0.086$, below the Planck 2018 bound $\Delta N_{\rm eff}\leqslant0.30$~\cite{Planck:2018vyg}. Along the $m_s<m_h$ WIMP curve, kinetic decoupling can occur well below the QCD phase transition in some regions, allowing $\Delta N_{\rm eff}$ to reach $\sim0.8$. However, those regions are already excluded by direct detection. Therefore, the $N_{\rm eff}$ constraints yield no additional exclusion of the surviving WIMP parameter space.

\section{Conclusion}\label{sec:conclusion}

In this work, we investigate the DM scenarios within the CC $SU(2)_X$ gauge theory, where the dark scalar resides in the triplet representation. EW symmetry breaking is triggered dynamically by the $SU(2)_X$ sector via the CW mechanism, thereby relaxing the hierarchy problem. The model yields two viable DM candidates: the conventional WIMP and supercooled DM, the latter emerging from the unique FOPT from the CC framework. The minimality of the CC theory allows the entire parameter space to be concisely represented in the two-dimensional plane $(m_X, m_s)$, offering a unified perspective on the DM phenomenology. In the light scalar regime ($m_s < m_h$), the WIMP is excluded by direct detection, but the supercooled DM scenario remains viable and can be probed via future GW and LLP searches. In the heavy scalar regime ($m_s > m_h$), the WIMP candidate can be probed through collider and GW experiments, whereas the supercooled DM can also be detected by GWs.

Our work can be further extended in several directions. First, more direct probes of supercooled DM beyond indirect sensitivity via the FOPT GWs or the scalar $s$ remain to be explored. Second, since the GW prediction depends sensitively on the FOPT dynamics, it could be refined by incorporating higher-order corrections and a more rigorous treatment of the vacuum tunneling rate. Third, because the post-FOPT reheating inevitably dilutes any preexisting relics, including the baryon asymmetry, an appropriate baryogenesis mechanism must be incorporated. A particularly attractive avenue is nonthermal leptogenesis, which can be realized either through bubble-assisted production or via the post-FOPT generation of heavy right-handed neutrinos. We defer a comprehensive exploration of this direction to future work.

\acknowledgments

We thank Felix Br\"{u}mmer, Giacomo Ferrante, and Shao-Ping Li for the helpful discussions and comments. This work is supported by the National Science Foundation of China under Grant No. 12305108.

\appendix
\addcontentsline{toc}{section}{Appendices}
\addtocontents{toc}{\protect\setcounter{tocdepth}{0}}
\setcounter{figure}{0}
\renewcommand{\thefigure}{A\arabic{figure}}
\section{Parametrization scheme}\label{app:parametrization}

Defining $B=3g_X^4/(4\pi^2)$, following Ref.~\cite{Liu:2024fly}, the model parameters are re-parametrized as
\be
\lambda_{hs}=\frac{w (m_s^2+m_h^2)\pm\sqrt{w^2 (m_h^2-m_s^2)^2-4 m_s^2 m_h^2 v^2}}{-2 w \left(v^2+w^2\right)}\, ,\ee
\be
B=\frac{(m_s^2+m_h^2) w\mp\sqrt{w^2 (m_h^2-m_s^2)^2-4 m_s^2 m_h^2 v^2}}{2w^3}\, ,
\ee
\be
\tan\theta=\frac{\pm2m_h^2v}{\pm(m_h^2-m_s^2)w+\sqrt{w^2 (m_h^2-m_s^2)^2-4 m_s^2 m_h^2 v^2}}\, ,
\ee
for $m_s\lessgtr m_h$, and $\lambda_h=-\lambda_{hs}w^2/(2v^2)$.

Since $w\gg v$, we can expand the fields around the vacuum to obtain the interactions between physical particles, i.e.,
\be
\mL_{3}\supset-\frac{m_h^2}{2v}h^3+\frac{m_h^2(2m_h^2+ m_s^2)}{2w( m_s^2- m_h^2)}h^2 s
+\frac{m_h^2 m_s^2 v(m_h^2-4 m_s^2)}{2 w^2 (m_h^2-m_s^2)^2}h s^2-\frac{5 m_s^2}{6 w}s^3\, ,
\ee
for triple couplings, and
\bea
\mL_{4}&\supset&-\frac{m_h^2}{8v^2}h^4
-\frac{m_h^4}{2}\left[\frac{(m_s^4-2 m_h^4 ) v}{(m_h^2 - m_s^2)^3 w^3} + \frac{1}{(m_h^2 - m_s^2) v w}\right]h^3s\\
&&+\frac{m_h^2}{4 w^2}\left[1-\frac{3 m_h^4}{(m_s^2-m_h^2)^2}\right]h^2s^2
+\frac{m_h^2 (5 m_h^4 m_s^2 - 19 m_h^2 m_s^4 + 11 m_s^6) v}{6 (m_h^2 - m_s^2)^3 w^3}hs^3-\frac{11 m_s^2 }{24 w^2}s^4\, ,\nn
\eea
for quartic couplings, regardless of the mass ordering between $h$ and $s$.

\setcounter{figure}{0}
\renewcommand{\thefigure}{B\arabic{figure}}
\section{Analytical estimates of ultra-supercooled FOPT dynamics}\label{app:PTDyna}

The scaling behavior of the FOPT parameters in the ultra-supercooling case admits a simple analytical understanding. The Euclidean action is given by
\be
\frac{S_3}{T} = \frac{1}{T} \int_0^\infty 4\pi r^2 \d r \left[ \frac{1}{2} \left( \frac{\d s_b}{\d r} \right)^2 + V_T(s_b(r), T) \right]\, ,
\ee
where $s_b(r)$ is the $O(3)$-symmetric bounce solution from the equation of motion
\be
\frac{\d^2 s_b}{\d r^2} + \frac{2}{r} \frac{\d s_b}{\d r} = \left. \frac{\d V_T(s, T)}{\d s} \right|_{s = s_b}\, ,
\ee
subject to the boundary conditions
\be
\left. \frac{\d s_b}{\d r} \right|_{r=0} = 0 \, , \quad \lim_{r \to \infty} s_b(r) = 0 \, .
\ee
In the regime of ultra-supercooling, the phase transition occurs at a temperature $T_*$ very close to $T_{\text{roll}}$. In this limit, the bounce solution is primarily determined by the structure of the potential barrier rather than the depth of the true vacuum.

\begin{figure}
\centering
\includegraphics[width=0.55\linewidth]{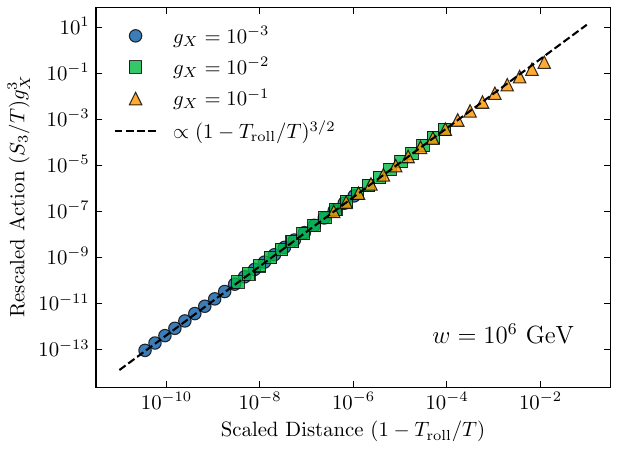}
\caption{Scaling of $(S_3/T)g_X^3$ with $(1 - T_{\text{roll}}/T)$. At fixed $w = 10^6$ GeV, numerical results for $g_X = 10^{-1}$, $10^{-2}$, and $10^{-3}$ collapse onto a universal line, confirming the analytical scaling of Eq.~\eqref{eq:S3_approx} (dashed line).}
\label{fig:figure1}
\end{figure}

Let $s_1$ denote the position of the local maximum of $V_T(s,T)$ separating the false vacuum at $s=0$ from the true vacuum basin, and let $V_1 \equiv V_T(s_1, T)$ be the corresponding barrier height. We further define $s_0$ as the field value where the potential vanishes immediately after the barrier, i.e., $V_T(s_0,T)=0$ for $s_0>s_1$. To analyze the scaling behavior, we rescale the field, potential, and radial coordinate as follows:
\be
\tilde{s} = \frac{s}{s_0}\, , \quad \tilde{V} = \frac{V_T(s, T)}{V_1}\, , \quad \tilde{r} = \sqrt{\frac{V_1}{s_0^2}} \, r \, . \label{eq:rescale}
\ee
In the CC $SU(2)_X$ model, $s_0\sim T / g_X$ is determined by the competition between the thermal mass ($\sim g_X T$) and the CW potential ($\sim g_X^2$). Remarkably, the barrier height then scales as $V_1 \sim T^4$, independent of $g_X$ up to an overall prefactor \cite{Iso:2017uuu}. With the above rescaling, the thermal action factorizes as:
\be
\frac{S_3}{T} = \frac{4\pi}{T} \frac{s_0^3}{\sqrt{V_1}} \tilde{S}_3 \equiv \frac{\mathcal{A}}{g_X^3} \tilde{S}_3(\tilde{T}) \, ,
\ee
where $\mathcal{A}$ is a coefficient and $\tilde{S}_3(\tilde{T})$ is a dimensionless shape factor. Near $T_{\rm roll}$, the shape factor could be evaluated by the thick-wall approximation, yielding~\cite{Linde:1981zj}
\be\label{eq:S3_approx}
\frac{S_3}{T} \propto \frac{1}{g_X^3} \left( 1 - \frac{T_{\rm roll}}{T} \right)^{3/2} \,,
\ee
as illustrated in Fig.~\ref{fig:figure1}.

For ultra-supercooled FOPT in the Type-I evolution pattern, the transition temperature $T_*\approx T_{\text{roll}}$, and the nucleation condition $\Gamma(T_n)/H^4(T_n)=1$ simplifies to
\be\label{eq:S_nuc}
\left.\frac{S_3}{T}\right|_{T_n} \approx 4\log\frac{M_{\rm Pl}\,m_h\,v_{\rm QCD}}{g_X^3 w^3}\,.
\ee
Over most of the parameter space, $T_n$ closely tracks the percolation temperature $T_*$, allowing us to use the nucleation criterion to extract scaling behavior. Equating the bounce action in \Eq{eq:S3_approx} with \Eq{eq:S_nuc} reveals that $(1-T_{\rm roll}/T_n)\sim g_X^2$ is required to compensate the prefactor $g_X^{-3}$ in the action, yielding
\be
\frac{\beta}{H} \sim \frac{1}{g_X^3} \frac{\d}{\d T}\left(1-\frac{T_{\rm roll}}{T_n}\right)^{3/2} \sim g_X^{-2}\,,
\ee
which accounts for the alignment of $g_X$ and $\beta/H$ contours observed in Fig.~\ref{fig:foptparam}.

The qualitative behavior of $R_0$ follows from the same rescaling argument used in \Eq{eq:rescale}. With $s_0 \sim T/g_X$ and $V_1 \sim T^4$, one obtains $R_0\sim 1/(g_X T)$. In the Type‑I evolution pattern, the ultra‑supercooled FOPT is extremely fast ($\beta/H\gg 1$), and bubble nucleation is sharply localized near the temperature $T_*$, which coincides with the rolling temperature $T_{\rm roll}$. Hence,  
\be
R_0 \sim \frac{1}{g_X T_{\rm roll}} \sim  \frac{w}{m_h v_{\rm QCD}} \, .
\ee
Thus, a stronger supercooling leads to a larger $R_0$, which can no longer be ignored in this regime.

\section{Monopole as the topological soliton dark matter candidate}\label{app:monopole}

The triplet $SU(2)_X$ model supports a topological soliton, known as the 't Hooft-Polyakov monopole~\cite{tHooft:1974kcl,Polyakov:1974ek}. For $m_X \gtrsim m_s$, the monopole has a radius $R_{\rm mp} \approx m_X^{-1}$ and a mass $M_{\rm mp} \approx 4\pi w / g_X$~\cite{Bai:2020ttp}. These monopoles are formed in the early Universe via the Kibble-Zurek mechanism when the $SU(2)_X \to U(1)_X$ symmetry breaking occurs~\cite{Kibble:1976sj,Zurek:1985qw}. In our CC model, this breaking is driven by a FOPT. Consequently, the number density of monopoles immediately after formation can be estimated from the density of true vacuum bubbles at $T_*$, parametrized as~\cite{Einhorn:1980ik,Yang:2022quy}
\be\label{Zurek}
n_{\rm mp}^*=pn_b^*\, ,
\ee
where $p \sim10^{-1}$, and $n_b^*$ is the bubble number density at $T_*$.

After production, monopoles and anti-monopoles may undergo annihilation, whose evolution is generally described by a set of Boltzmann equations~\cite{Preskill:1979zi}. However, if no dark‑photon thermal bath exists following the FOPT, annihilation processes are strongly suppressed, and the monopoles simply free‑stream without significant interaction~\cite{Bai:2020ttp}. In this case, their number density is solely diluted by the entropy production during reheating, and they scale as cold DM thereafter. The present‑day relic abundance is then given by
\be
\Omega_{\rm mp}h^2=\frac{n_{\rm mp}^*}{s_*}\frac{T_{\rm rh}}{T_\Lambda}\left(\frac{T_*}{T_\Lambda}\right)^3s_0M_{\rm mp}\frac{8\pi }{3M_{\rm Pl}^2}\left(\frac{h}{H_0}\right)^2\, ,
\ee
where $s_*=2\pi^2g_{*s}T_*^3/45$ is the entropy density at FOPT. In the parameter space of Fig.~\ref{fig:DM}, the monopole relic abundance is always subdominant to that of $X^\pm$, and therefore monopoles cannot be DM candidates, consistent with Ref.~\cite{Brummer:2025inh}.

\section{Definition of $\ave{\sigma v_{\rm rel}}$}\label{app:xsection}

For a Maxwell‑Boltzmann distribution, the thermally averaged cross section for the process $ab\to cd$ is~\cite{Gondolo:1990dk}
\begin{multline}
\ave{\sigma_{ab\to cd}v_{\rm vel}}=\frac{(8m_a^2m_b^2T)^{-1}}{K_2(m_a/T)K_2(m_b/T)}\int\d \hat s\sigma_{ab\to cd}(\hat s)\\
\times\left[1-\frac{(m_a+m_b)^2}{\hat s}\right]\left[1-\frac{(m_a-m_b)^2}{\hat s}\right]\hat s^{3/2}K_1\left(\frac{\sqrt{\hat s}}{T}\right),
\end{multline}
where $\sigma_{ab\to cd}(\hat s)$ is the initial-state-averaged cross section, and $\hat s$ is the square of the center‑of‑mass energy. The dominant reaction channels for $X^+X^-$ annihilation in our study are listed in \Eq{ann}, i.e.,
$\ave{\sigma v_{\rm vel}}\equiv \ave{\sigma_{X^+X^-\to ss}v_{\rm vel}}+\ave{\sigma_{X^+X^-\to {\rm SM\,SM}}v_{\rm vel}}$. Besides, we also consider the following processes
\be
ss/sh\to hh/t\bar t/W^+W^-/ZZ\,;\quad sZ\to W^+W^-\,,
\ee
and all their crossed processes to verify that $s$ can thermalize with the SM plasma after the FOPT.

Since the $X^\pm$ bosons are charged under the residual $U(1)_X$, they experience a long‑range attractive force mediated by the massless dark photon $A'$. This interaction modifies the non‑relativistic annihilation cross section through the Sommerfeld enhancement. To incorporate this effect, we multiply $\ave{\sigma v_{\rm vel}}$ by an extra factor $\mS$~\cite{Arkani-Hamed:2008hhe}
\be
\mS=\frac{\pi\alpha_X/v_{\rm dm}}{1-e^{-\pi\alpha_X/v_{\rm dm}}}\,,
\ee
with $\alpha_X=g_X^2/(4\pi)$ and $v_{\rm dm}\approx\sqrt{6/z}$. Although the Sommerfeld factor modestly increases the annihilation rate, it remains close to unity in the parameter region of interest due to the small values of $g_X$.

\addtocontents{toc}{\protect\setcounter{tocdepth}{3}}

\bibliographystyle{JHEP-2-2.bst}
\bibliography{references}

\end{document}